\documentclass[prd,amsfonts,twocolumn,nofootinbib,showpacs]{revtex4}
\RequirePackage{graphicx}
\usepackage{enumerate}

\begin{document}
\title{Particle production with L-R neutrino oscillation} 

\author{Seishi Enomoto}
\affiliation{Institute of Theoretical Physics, Faculty of Physics,
University of Warsaw, Ho$\dot{z}$a 69, 00-681 Warsaw, Poland}
\author{Tomohiro Matsuda}
\affiliation{Laboratory of Physics, Saitama Institute of Technology, Fukaya, Saitama 369-0293, Japan}
\begin{abstract}
When the Higgs field starts oscillation after Higgs inflation,
gauge bosons are produced non-perturbatively near the Enhanced
 Symmetry Point (ESP).  
Just after the particle production, when the Higgs field is going away
 from the ESP, these gauge bosons gain mass and decay or annihilate into
 Standard Model (SM) fermions. 
Left-handed neutrinos can be generated in that way.
If one assumes the see-saw mechanism, the mass matrix of a pair
 of left and right-handed neutrinos is non-diagonal.
Although their mixing in the mass eigenstates is negligible in the
 true vacuum, it could be significant near the edge of the Higgs
 oscillation, where the off-diagonal component is large.
Therefore, the left-handed neutrinos generated from the gauge bosons
can start neutrino oscillation between the right-handed neutrinos.
We study the particle production when such L-R neutrino oscillation is significant.
For a working example, the non-thermal leptogenesis scenario after Higgs
 inflation is examined, which cannot be realized without the L-R
 neutrino oscillation. 
The same mechanism could be applied to other singlet particles
whose abundance has been neglected.
\end{abstract}

\maketitle

\section{Introduction and the model}
The recent release of the Planck data~\cite{Ade:2015lrj} suggests that
the Higgs inflation model~\cite{Bezrukov:2007ep, Bezrukov:2008ut} is in agreement with
the Planck constraints.\footnote{The scale-dependence of the spectrum
can be shifted by an additional degree of freedom~\cite{Kohri:2014jma}.
Observation of the tensor mode and its spectral index (if possible) could fix such
ambiguity~\cite{Kohri:2015laa}.}  
The idea of Higgs inflation is quite attractive since it uses the well-known scalar
field and explains inflation without introducing a new field to the SM.
On the other hand, it is hard to explain the
baryon asymmetry of the current Universe within the SM.
The neutrino mass, which is discovered to be non-zero in
1998~\cite{Fukuda:1998mi}, is also beyond the SM. 
One way to explain the neutrino mass is to consider see-saw
mechanism~\cite{Yanagida:1979as}, which is also expected to explain the
baryon asymmetry via leptogenesis~\cite{Fukugita:1986hr}.
In this paper we study a specific process of generating right-handed
neutrinos during non-perturbative particle production (preheating) 
that may occur before reheating.
We apply the idea of neutrino oscillation to the mass matrix of the
see-saw mechanism.
This idea works when the Higgs field has large
vacuum condensate.
Showing that significant mixing is possible during preheating,
we demonstrate how the L-R neutrino oscillation works to generate
right-handed neutrinos. 
Remember that conventional neutrino oscillation in the dense
matter~\cite{Mikheev:1986wj, Wolfenstein:1977ue} leads to a similar
situation.
In that case the matrix is given for a pair of $\nu_e$ and $\nu_\mu$.
The diagonal element of $\nu_e$ becomes time-dependent due to the
interaction between matter, while the other elements remain constant.
In the core of the star, where the density is high, the mass eigenstates
could be $\psi_+\simeq \nu_e$ (heavy state) and $\psi_-\simeq \nu_\mu$
(light state), while outside the star
(in the vacuum) these relations turn out to be opposite.
If $\nu_e\simeq \psi_+$ is produced in the core and it propagates
adiabatically until it reaches the surface of the star, one will find emission of
the heavier mass eigenstate $\psi_+$ from the star, which is now $\psi_+\simeq \nu_\mu$. 
In our case, the left-handed neutrino ($\nu_L$) is generated near the edge, where 
$\nu_L$ is a mixed state of $\psi_+$ and $\psi_-$.
Of course, one will find $\psi_+\simeq N_R$ in the true vacuum.
Although our situation could be unusual, the use of the time-dependent
matrix for the neutrino production is quite common in
astro-particle physics. 
We apply our idea to the simplest leptogenesis scenario and find
that the new mechanism assists the non-thermal leptogenesis scenario.

\subsection{Preheating and neutrino oscillation after Higgs inflation}
Since in this paper we have no space to review the details of the
oscillation and the particle 
production after Higgs inflation, we carefully follow Ref.\cite{GarciaBellido:2008ab,
Bezrukov:2008ut} to avoid confusions. 
Here we prepared our starting point as simple as possible.
We start with the effective potential
\begin{eqnarray}
\label{eff-pot}
V(\chi)&=&\frac{1}{2}M^2\chi^2,
\end{eqnarray}
where $M\equiv \sqrt{\lambda}M_p/\sqrt{3}\xi$ and
$\chi$ comes from the original Higgs field after conformal
transformation.
Here $\lambda$ denotes the original quartic coupling of the Higgs field.
The conformal factor of the transformation can be expressed using 
\begin{eqnarray}
\Omega^2(h)&=&1+\frac{\xi h^2}{M_p^2}\\
\Omega^2(\chi)&=&e^{\alpha \kappa \chi}.
\end{eqnarray}
We used $\alpha\equiv\sqrt{2/3}$ and $\kappa\equiv M_p^{-1}$,
where $M_p$ is the reduced Planck mass of the original action.
Here $h$ denotes the condensate of the original Higgs field in the
unitary gauge. 
The oscillation starts when $X_0 \sim 0.1 M_p$, where $X_n$ denotes the
amplitude of the $n$-th oscillation.
Consider the masses of ``gauge bosons'' in the Einstein frame;
\begin{eqnarray}
{\cal L}^{E}_{W,Z}&\sim& \tilde{m}_W^2 \tilde{W}_\mu^{+} 
\tilde{W}^{\mu-}
+\frac{1}{2}\tilde{m}_Z^2 \tilde{Z}_\mu \tilde{Z}^{\mu},
\end{eqnarray}
where 
\begin{eqnarray}
\tilde{m}^2_W&\equiv&\frac{g^2_2 M_p^2(1-e^{-\alpha\kappa|\chi|})}{4\xi}\\
\tilde{m}_Z^2&\equiv& \frac{\tilde{m}_W^2}{\cos \theta_W}.
\end{eqnarray}
Here $g_2$ is the $SU(2)_L$ gauge coupling constant at the corresponding
scale and $\theta_W$ is the Weinberg angle defined at the scale.
Tilde denotes quantities in the Einstein frame.

Similarly, the interaction between fermions and gauge bosons can be
expressed as 
\begin{eqnarray}
&\sim& 
\frac{g_2}{\sqrt{2}}\tilde{W}_\mu^{+} \tilde{J}^{\mu-}
+\frac{g_2}{\sqrt{2}} \tilde{W}^{-}_\mu\tilde{J}^{\mu+}
+\frac{g_2}{\cos\theta_W} \tilde{Z}_\mu \tilde{J}^{\mu0},
\end{eqnarray}
where $\tilde{J}^{\mu(\pm,0)}$ are charged and neutral currents of the SM
fermions, which may contain left-handed neutrinos.

For a given family of the quark sector the mass of the fermions ($f$ :
the family index is omitted) is
found from the Yukawa coupling. In the Einstein frame it can be written
in a schematic form 
$\tilde{m}_f \bar{\tilde{f}}\tilde{f}$, where 
\begin{eqnarray}
\tilde{m}_f&\equiv&\frac{y_f
 M_p}{\sqrt{2\xi}}\left(1-e^{-\alpha\kappa|\chi|}\right)^{1/2}.
\end{eqnarray}

On the other hand, the neutrino mass is given by the see-saw mechanism.
We choose the mass matrix for the left ($\nu_L$) and the 
right ($N_R$) handed neutrinos;
\begin{eqnarray}
{\cal M}=\left[
\begin{array}{cc}
0 &m_D\\
m_D &M_R
\end{array}
\right],
\end{eqnarray}
where we take $m_D=y h$.\footnote{Here we consider only the first generation for sake of simplicity.
In reality the flavor structure is very important. It is well known that in thermal leptogenesis
the evolution of the lepton asymmetry depends on the flavor structure.
We need to introduce more than two right-handed neutrinos to explain the observed active neutrino mass differences.
Moreover, CP violation requires at least three generation.}
${\cal M}$ can be diagonalized using the rotation matrix $U(\theta)$ as
${\cal M}_\mathrm{diag}=U^{T}{\cal M}U\equiv diag(M_-
,M_+)$ to give\footnote{See also Ref.\cite{Drewes:2013gca}.}
\begin{eqnarray}
\label{eq-mass-matrix}
\frac{1}{2} \left(
\begin{array}{cc}
M_R-\sqrt{M_R^2+4m_D^2} & 0\\
0 & M_R+\sqrt{M_R^2+4m_D^2} 
\end{array}\right),
\end{eqnarray}
where the angle of the rotation is given by
\begin{eqnarray}
\theta &=& \frac{1}{2}\tan^{-1}\left(\frac{2m_D}{M_R}\right).
\end{eqnarray}
Conformal transformation leads to
\begin{eqnarray}
\tilde{M}_R^2&=& M_R^2\Omega^{-2}=M_R^2 e^{-\alpha\kappa
|\chi|}\\
\tilde{m}_D^2 &=& \frac{y_{\nu}^2
M_p^2(1-e^{-\alpha \kappa |\chi|})}{\xi}.
\end{eqnarray}
Practically, these variables are simplified using $1-e^{-\alpha \kappa
|\chi|}\simeq \alpha \kappa |\chi|$.

We consider the simplest see-saw mechanism in which heavy $N_R$
($M_R\gg h$ in the vacuum)
 is responsible for the SM neutrino mass.

In terms of the mass eigenstates ($\psi_\mp$), left and right-handed
neutrinos are written as
\begin{eqnarray}
\nu_L&=&\psi_{-} \cos \theta -\psi_{+} \sin\theta \\
N_R&=&\psi_{-} \sin \theta +\psi_{+} \cos\theta.
\end{eqnarray}
Normally (in the vacuum) one will find $M_R\gg m_D$ that leads to 
$\theta\simeq m_D/M_R\simeq 0$.
Therefore there is no significant left-right mixing in the see-saw
sector of the vacuum state.
On the other hand, although temporarily the mixing could be significant
when $m_D$ is as large as $M_R$.

In Ref.\cite{Nilles:2001fg}, one can find extensive
study of the non-perturbative particle production due to the time-dependent
non-diagonal mass matrix (i.e, when $\dot{U}\ne 0$).
Ref.\cite{Enomoto:2014cna} studies a model in which the mass
term is not time-dependent but the interaction between other fields
causes particle production. 
Alternatively one may consider higher dimensional
interaction~\cite{Bezrukov:2008ut, Enomoto:2013mla, Enomoto:2014hza}.
However, direct production of fermions is 
not important in our case since fermions do not lead to parametric
resonance.\footnote{Notice also that conventional fermion 
preheating~\cite{Giudice:1999fb} uses  
$m_f=M_0+g\phi$ that vanishes at $\phi=-M_0/g$, while the mass
of $\psi_+$ cannot vanish in the see-saw mechanism.}

Besides neutrinos, other particles such as gauge bosons and
quarks can be generated non-perturbatively.
Among those, the most significant effect is expected for the gauge
bosons because of the parametric resonance.
However, since the gauge bosons may have huge mass when the amplitude of
the oscillation is large, they can partially decay into fermions before
the next particle production.
This gives suppression of the resonance in the first stage.
Then the gauge bosons accumulate slowly and finally the parametric resonance
takes place.
According to Ref.\cite{Bezrukov:2008ut}, the main process responsible
for the energy transfer is the annihilation of the W bosons.
Then the radiation starts to dominate when the amplitude is $X_r\sim M$.
Thermalization takes place later at $X_R\le X_r$.

Since the currents that couple to those gauge bosons contain neutrinos,
SM neutrinos can be generated when gauge bosons decay or annihilate.
{\bf One thing that is unusual here is that the $\nu_L$-state, which is
produced from the gauge bosons, can be a mixed state of the mass
eigenstates $\psi_\pm$ if the transfer proceeds near the
edge.}\footnote{Consider  
expansion of $h$ around the expectation value of 
 the condensate; $h \rightarrow v + \hat{h}$.
Expanding the original interaction ($yhN_R\nu_L$) using mass eigenstates
 ($\psi_\pm$), one will find that the coefficient of the interaction
 $\hat{h}\psi_+  \psi_-$ is proportional to $\cos 2\theta$.
On the other hand, if $\psi_+$ does not decay during the oscillation, it
can be translated into $N_R$ (adiabatic process). 
See also Ref.\cite{Mikheev:1986wj, Wolfenstein:1977ue}.}
The condition $\tilde{m}_D>\tilde{M}_R$ can be satisfied until the amplitude decreases
to reach $X_e\simeq \xi M_R^2/(y_\nu^2M_p)\sim M_R^2/(y^2_\nu M)$.
Using $m_\nu\equiv y_\nu^2 v_{EW}^2/M_R$, we can write 
\begin{eqnarray}
X_e&\simeq&
 10^{14}\left(\frac{M_R}{M}\right)\left(\frac{v_{EW}}{10^{2} \mathrm{GeV}}\right)^2
\left(\frac{0.1 eV }{m_\nu}\right)\mathrm{GeV}.
\end{eqnarray} 

Here we can assume that neutrinos generated from the gauge bosons are
relativistic.(See Ref.\cite{Bezrukov:2008ut}.)
Therefore, typical time scale of the neutrino oscillation is measured by the
square mass difference between $\psi_+$ and $\psi_-$\footnote{See
Eq.(\ref{eq-mass-matrix}).}.
This gives the simple estimation of the
typical time scale $\sim \frac{\Delta M_\pm^2}{4E}\sim
\frac{M_R\sqrt{M_R^2+4m_D^2}}{m_D}\sim M_R$, which is identical to the
right-handed neutrino mass.

The most obvious realization of the scenario is to take $M_R >M$.
In that case the neutrino oscillation is fast enough to make 
oscillation before the Higgs field goes back to the origin.
If one compares our result with Ref.\cite{Bezrukov:2008ut}, one will
find that tiny Yukawa couplings are not important in our case, but small 
$M_R$ makes the oscillation length large and may prevent
      L-R oscillations.
In this paper we focus on this simplest scenario and examine the non-thermal
leptogenesis scenario.

Although we are not discussing the possibility in this paper, less
obvious scenario 
could be possible when $M_R \sim H$, where $H$ is the 
Hubble parameter during oscillation.
In this case, the ``amplitude'' of the Higgs oscillation is large enough
to keep the maximum mixing, and the time scale of the neutrino
oscillation is comparable to the time scale of the cosmological
evolution of the amplitude.
Indeed, coarse graining the Higgs oscillation and consider the amplitude
of the oscillation as the time-dependent parameter of the model,
one could be able to find significant
production of right-handed neutrinos when $M_R >H$.
Since in this case the typical time scale of the L-R neutrino
oscillation is not very short, one has to consider the quantum Zeno effect.
In the hot and dense environment (when the scattering of SM neutrinos is
      significant) there is a competition between the oscillation length
      and the mean free path.
If the oscillation length is much larger than the mean free path, 
the L-R transition probability is hindered.
This is the quantum Zeno effect.
Usually one has to introduce a damping term in the neutrino oscillation
equations~\cite{Boyanovsky:2006it}.
However, in the complicated far-from equilibrium environment of
preheating it is quite difficult to quantify the damping.\footnote{We
are focusing on changes in the initiatial condition. 
Therefore, active-sterile neutrino oscillation  
after reheating is not considered here.} 
We are going to reserve such issue for future study.

Besides that, it should be noted that partial translation could be
important even if the neutrino oscillation is blocked halfway,
since the L-R translation is much more efficient compared with the usual
process.
Importantly, the same mechanism can be applied to other singlet particles 
whose abundance has been neglected.

Eventually, gauge bosons can produce $N_R$ via the unconventional
``neutrino oscillation'' and the production mechanism can affect
cosmological scenarios after Higgs inflation.

\subsection{Non-thermal Leptogenesis from the L-R neutrino oscillation}

The main purpose of this paper is to present a novel mechanism
of generating particles from a time-dependent off-diagonal element in the
mass matrix.
The mechanism will work in general situation.
Indeed, the mechanism may work in Grand
Unified Theory (GUT), which will have many non-diagonal mass matrices.
The calculation is quite complicated and the study
requires careful numerical calculation.

Nevertheless, we believe that our attempt to show an
application of the mechanism in the simplest set-up is very useful, 
as far as it gives an intuitive estimation of the quantities under some 
reasonable assumptions.
Among possible applications of the model, dark matter production and
leptogenesis would be 
the most important.
In this paper, we choose the simplest model of non-thermal
leptogenesis in which the non-equilibrium decay of the right-handed
neutrino generates baryon (lepton) asymmetry of the Universe.
In our model, the right-handed neutrino is never thermalized and there is
no wash-out of the asymmetry.
This is our simplest set-up of the model.
The leptogenesis proceeds when $\psi_+$ decays into $\chi$ 
and $\psi_-$, where  $\psi_+$ is a mixed state of $N_R$ and $\nu_L$. 
$\psi_-$ turns into $\nu_L$ in the vacuum.

According to Ref.\cite{GarciaBellido:2008ab, Bezrukov:2008ut},
$\chi$ starts oscillation at $X_0\sim 0.1 M_p$ and parametric resonance is
(partially) suppressed until $X_{PR}\sim 10^{15}$GeV.
At $X_{PR}$, we expect that 1/10 of the
inflaton energy is translated into gauge bosons.
This estimation is taken from Ref.~\cite{GarciaBellido:2008ab}.
Therefore we assume
\begin{eqnarray}
\label{rhoB01} 
\rho_{Bs}(t_{PR})&\simeq& 0.1 \rho_\chi(t_{PR}).
\end{eqnarray}
Then we assume that the energy transfer is accomplished and
thermalized when
$X_{R}\sim X_r\sim M$.
This gives the reheating temperature $T_R\sim 10^{13}$ GeV.\footnote{As
is noted in Ref.\cite{GarciaBellido:2008ab}, estimation of the
reheating temperature is not simple. Here we took $T_R$ 
from Ref.\cite{Bezrukov:2008ut}.} 
During the process, certain fraction of gauge bosons is continuously
converted into fermions. 
In reality the fraction is time-dependent and the calculation of the number
densities is highly non-local (i.e, they must be calculated as the time
integral of a complex system).
To avoid further complexity, we assume $X_{PR}\simeq X_{e}$ and 
define a new parameter $\epsilon_{+}<1$, assuming that the number
density of $\psi_+$ just after $X_{PR}\simeq X_e$ is given by
\begin{eqnarray}
n_+&\equiv& \epsilon_+ n_{Bs},
\end{eqnarray}
 where $n_{Bs}\equiv n_Z+n_W$ is the number density of the gauge bosons.
Since $\psi_+$ can decay into the Higgs and $\psi_-$, 
we can define $\epsilon_{CP}$ to estimate the produced lepton number 
\begin{eqnarray}
n_L&\sim& \epsilon_{CP}n_+\\
&\sim&  \epsilon_{CP} \epsilon_{+} \frac{0.1M^2 X_{PR}^2}{g_2 \sqrt{M X_{PR}}}\\
& \sim& \epsilon_{CP} \epsilon_{+} \left(MX_{PR}\right)^{3/2},
\end{eqnarray} 
where $m_Z\sim m_W\sim g_2 \sqrt{MX}$ is assumed for simplicity.
In the vacuum, when the background is static, $\epsilon_{CP}$ is
identical to the conventional leptogenesis scenario, while in the above
case one has to calculate the integral of the system to obtain 
$\epsilon_{CP}$.
In that sense both $\epsilon_+$ and $\epsilon_{CP}$ are not local.
Actual computation of those parameters is very difficult, and it
depends on the model parameters of the lepton sector.\footnote{See
also Ref.\cite{Asaka:2005an, Canetti:2012kh, Drewes:2013gca}, in which
L-R neutrino oscillation after reheating is considered for smaller
$M_R$.}
Nevertheless, speculation would be possible for these parameters, which
will be helpful for our purpose in this section.
Since the decay or the annihilation before $t_{PR}$ is sufficient to prevent the
resonance, $\epsilon_+\gtrsim 0.01-0.001$ would be
conceivable.\footnote{Here $\epsilon_+$ is not identical to the
branching ratio of the gauge boson. At the beginning of the oscillation
the decay is sufficiently fast near the edge and almost all $n_{B_s}$ is
transmitted to the decay products.
If we introduce the ratio $\epsilon_d$ that measures the efficiency of
the decay, we can write $\epsilon_+=\epsilon_{br}\epsilon_d$, where
$\epsilon_{br}$ denotes the branching ratio.
At first $\epsilon_+$ is almost identical to the branching ratio but
during the oscillation $\epsilon_d$ decreases with time.}
If the conventional CP violation is valid for our case,
one can expect $\epsilon_{CP}\sim 10^{-5}$, which can be enhanced.
Here the family multiplicity is implicitly assumed for $N_R$.
If we can assume $n_L\propto a(t)^{-3}$ and $\rho_\chi\propto a(t)^{-3}$
between $X_e$ and $X_r$, where
$a(t)$ is the scale factor of the Universe, we find using
$n_{L}(t_{PR})=\epsilon_{CP}\epsilon_+n_{B_s}=
\epsilon_{CP}\epsilon_+\rho_B/m_B=0.1 \times
\epsilon_{CP}\epsilon_+\rho_\chi/m_B$ and 
$\rho_\chi(T_r)\simeq \rho_\chi(T_R)\simeq 10^2 T_R^4$; 
\begin{eqnarray}
\frac{n_L(t_R)}{T_R^3}&\sim&
0.1 \times \epsilon_{CP}\epsilon_{+}
\frac{n_L(t_R)}{n_L(t_{PR})}\frac{\rho_\chi(t_{PR})}{T_R^4}
\frac{T_R}{\tilde{m}_B}\\
&\sim& 10\times \epsilon_{CP}\epsilon_{+}
\frac{T_R}{g_2 \sqrt{MX_{PR}}}\\
&\sim&\epsilon_{CP} \epsilon_{+}.
\end{eqnarray}
Here we used $n_L(t_R)/n_L(t_{PR})=\rho_\chi(t_{R})/\rho_\chi(t_{PR})$.
In reality the above estimation could be enhanced since $n_L$ is
supplied by the gauge bosons while 
$\rho_\chi$ decreases more rapidly due to the decay.
 
Alternatively, the number density of the gauge bosons produced at the ESP can be estimated
by using so-called instant preheating~\cite{Felder:1998vq}.
The simplest way discussed in Ref.\cite{Felder:1998vq} is to estimate
the time for the instantaneous particle production $\Delta t_* \sim
\chi_a /\dot{\chi}$ and use this to obtain the occupation number 
$n_k=\exp(-\pi k^2/k_*^2)$, where $k_*\sim \Delta t_*^{-1}$.
\footnote{The power of $k$ in the above formula $n_k$ is correct in the
standard preheating scenario, where the mass depends on the
oscillating field as $m^2\propto \chi^2$.
Since in the present scenario we are considering $m^2\propto \chi$,
there could be some deviation in $n_k$.
See Ref.\cite{Enomoto:2013mla, Enomoto:2014hza} for more details.
This can change the numerical factor in the final result.}
Here$|\chi|<\chi_a$ denotes the region in which the adiabatic condition
is violated.
In the present case we find~\cite{GarciaBellido:2008ab} $\chi_a\simeq
\left(\frac{\xi |\dot{\chi}|^2}{\alpha g^2M_p}\right)^{1/3}$ and 
\begin{eqnarray}
\label{pr-or}
n_{B_s}(t_n)&=& \frac{1}{2\pi^2}\int^{\infty}_0dk k^2 n_k\nonumber\\
&\simeq&\frac{k_*^3}{8\pi^3}\nonumber\\
&=& \frac{\lambda g^2 M_p \dot{\chi}}{8\pi^3 \xi}\sim {\cal O}(10^{-3})\times g^2M^2 \chi_n,
\end{eqnarray}
where $n_{B_s}=0$ is assumed just before $\chi$ passes the ESP.
The above calculation is very crude but in good agreement with Eq.(89) in
Ref.\cite{GarciaBellido:2008ab}.
Now we have 
\begin{eqnarray}
\frac{n_+(t_R)}{T_R^3}
&\sim&\epsilon_{+} 10^{-3}\times\sqrt{\frac{M}{\chi_n}}.
\end{eqnarray}
Note that even in the above modest estimation the amount of the
right-handed neutrinos is not negligible.
This result gives a very modest estimation of the lepton asymmetry
\begin{eqnarray}
\frac{n_L(t_R)}{T_R^3}
&\sim&\epsilon_{+}\epsilon_{CP} 10^{-3},
\end{eqnarray}
which is still enough to realize non-thermal leptogenesis.

Before closing our discussion about leptogenesis, we have to make some
comments about the hierarchy and the naturalness between the electroweak
scale and $M_{R}$.
The quantum correction to the electroweak scale cannot meet the
naturalness criteria when $M_{R}$ is as large as $10^{14}$ GeV. 
If one demands supersymmetry, the particle production played by the
Higgs oscillation will be very different, since in that case there
could be many directions that can break the gauge symmetry at the same
time when the Higgs oscillates.

Although we calculated quantities in the simplest situation,
our result is illuminating the possibility of generating baryon(lepton)
number asymmetry during the preheating stage after Higgs inflation.
In contrast to thermal leptogenesis the process works for $M_R\ge
T_R$.

\section{Conclusion and discussion}
\label{conclusion}
Particle production is a very old idea.
Many scenarios have been studied to explain experiments and observations.
Among those, neutrino oscillation is a rather new idea, which is now
essential for the physics in the lepton sector of the SM.
Non-zero neutrino mass is needed for the neutrino oscillation, but it
has to be (unnaturally) light compared with other SM particles.
One way to explain the neutrino mass is to introduce the see-saw
mechanism, which extends the neutrino sector by adding a missing
right-handed partner of the left-handed neutrino.
If the see-saw mechanism works in the neutrino sector, Higgs oscillation
after Higgs inflation may lead to ``another neutrino oscillation''
between left and right-handed neutrinos.
A strange consequence of the L-R oscillation is that the sterile 
neutrino (right-handed neutrino) is generated from the gauge
bosons.
The process is very simple.
If the gauge bosons are transferred into SM-neutrinos when the mixing is
significant, the SM-neutrinos can be converted into right-handed neutrinos
via the left-right oscillation.
The process is obviously different from the so-called off-diagonal
preheating considered in past
studies~\cite{Nilles:2001fg}.\footnote{After integrating out the heavy
modes one will find higher dimensional interaction, which has been
considered for preheating in Ref.~\cite{Enomoto:2014hza}.}
In this paper we applied the idea to the simplest model of Higgs
inflation and found that leptogenesis is successful even if the
reheating temperature is lower than the right-handed neutrino mass.
At this moment our estimation is quite rough, since the particle
production occurs during oscillation and important parameters are
not defined local.
Moreover, the process starts with the non-perturbative production of
unstable gauge bosons, which may cause significant backreaction.
Therefore the actual calculation is quite complicated and very careful
numerical analysis is needed, especially when the flavor structure is
introduced.
For simplicity we used previous results in Ref.\cite{Bezrukov:2008ut,
GarciaBellido:2008ab} in our estimation and calculated the abundance of
$N_R$ from instant translation.
Nevertheless, it is quite obvious that for singlet fermions the process
of neutrino oscillation is
much more efficient compared with the usual process of particle
production \cite{Bezrukov:2008ut}.
Therefore, we conclude that the new mechanism we have considered in this
paper is very important for the physics related to the 
neutrino sector.

\section*{Acknowledgement}
S.E. is supported in part by the Polish NCN grant DEC-2012/04/A/ST2/00099.
T.M. would like to thank organizers and participants of COSMO14 where this
idea initiated.

\end{document}